\documentclass[a4paper,10pt,twoside]{cpc-hepnp}

\usepackage{multicol}
\usepackage{graphicx}
\usepackage{booktabs}
\usepackage{amssymb,bm,mathrsfs,bbm,amscd}
\usepackage[tbtags]{amsmath}
\usepackage{lastpage}
\usepackage{CJK}
\usepackage{epsfig}

\begin{document}
\begin{CJK*}{GBK}{song}


\fancyhead[r]{Submitted to Chinese Physics C }

\title{Uncoupled achromatic condition of a dog-leg system with the presence of RF cavities\thanks{Supported by National Natural Science Foundation of China (10875099, 91126003), the China ADS Project (XDA03020000)}}

\author{%
      GENG Hui-Ping\email{genghp@mail.ihep.ac.cn}%
\quad GUO Zhen
}
\maketitle

\address{%
Institute of High Energy Physics, Chinese Academy of Sciences, Beijing 100049, China\\
}

\begin{abstract}
To merge the beam from either of the two injectors to the main linac, a dog-leg system will be employed in the second Medium Energy Beam Transport~(MEBT2) line of the China ADS driving accelerator. The achromatic condition has to be guaranteed to avoid beam center excursion against energy jitter. RF cavities were found indispensable to control the bunch length growth in the dog-leg system of MEBT2. The full uncoupling between transverse and longitudinal plane is desired to minimize
the growth of projected rms emittances. The uncoupled achromatic condition of this dogleg system with the presence of RF bunching cavities will be deduced using the method of transfer matrixes. It is found that to fulfil the uncoupling condition, the distance between the bunching cavities is uniquely determined by the maximum energy gain of the RF cavities. The theoretical analysis is verified by the simulation code TraceWin. The space charge effect on the uncoupled achromatic condition and the beam emittance growth will also be discussed.
\end{abstract}

\begin{keyword}
dog-leg system, achromatic, coupling, transfer matrix, medium energy
\end{keyword}

\begin{pacs}
41.75.-i, 41.85.-p, 29.27.-a
\end{pacs}

\begin{multicols}{2}

\section{Introduction}

The C-ADS accelerator is a CW proton linac which uses superconducting
acceleration structures except the RFQs~\cite{cads}. To ensure the high reliability of
the accelerator at low energy, two parallel injectors are to be designed as
one will be 'hot spare' of the other.
the second Medium Energy Beam Transport (MEBT2)
is to transport and match the beam from either of the two injectors to the main linac.
A dog-leg system will be employed in the MEBT2
line of the China ADS accelerator.

Achromatic condition has to be guaranteed to
avoid beam center excursion against energy jitter. A latest design of MEBT2~\cite{mebt2} show
that RF cavities are indispensable to control the bunch length growth and emittance growth
in the dog-leg system of MEBT2. With the presence of RF cavities in a bending section, there will be
coupling between transverse and longitudinal planes which may cause
extra emittance growth. The full uncoupling between transverse and longitudinal planes is desired to minimize
the growth of projected rms emittance.

In this paper, the uncoupled achromatic condition of a dogleg system with the presence of RF bunching cavities
will be deduced using the method of transfer matrixes. The analytical result
will be checked by simulations carried out with the code TraceWin~\cite{tracewin}.
The space charge effect will also be briefly discussed.
\begin{center}
\includegraphics[width=8cm]{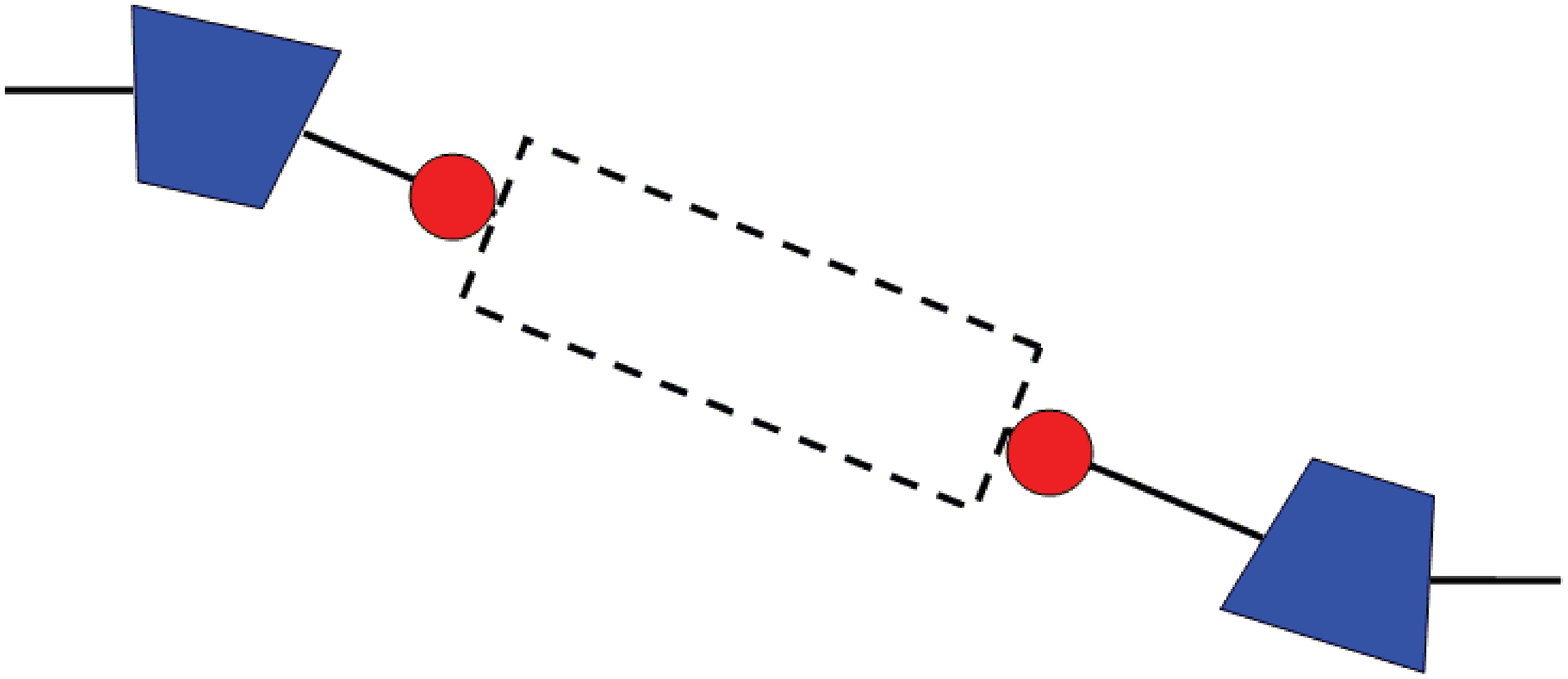}
\figcaption{\label{bend}   A typical layout of a dog-leg section with RF cavities.}
\end{center}

\section{Theoretical analysis}
The typical layout of a dog-leg section with the presence of two bunching cavities is shown in Fig.~\ref{bend}.
The two bending magnets bend the beam by the same angle but to opposite direction,  the red dot represent a
RF cavities and the dotted rectangle represents any combination of quadrupoles and drifts.

The transfer matrix of a bending magnet which bends the beam to the right can be written as,
\begin{eqnarray}
M_{b_+}=\left( \begin{array}{cccccc}
 b_{11} & b_{12} & 0  & 0    & 0   & b_{16}  \\
 b_{21} & b_{22} & 0  & 0    & 0   & b_{26}   \\
  0     & 0      & 1  & b_{34}&0   &  0     \\
  0     & 0      & 0  & 1    & 0   &  0     \\
-b_{26} & -b_{16}& 0  & 0    & 1   & b_{56}  \\
  0     & 0      & 0  & 0    & 0   & 1
\end{array} \right) .
\end{eqnarray}

A bunching cavity can be replaced by a RF gap for simplicity, thus,
its transfer matrix can be written as,
\begin{eqnarray}
M_{gap}=\left( \begin{array}{cccccc}
 1      & 0    & 0   & 0     & 0    & 0  \\
1/f_x   & 1    & 0   & 0     & 0    & 0   \\
  0     & 0    & 1   & 0     &0     &  0     \\
  0     & 0    & 1/f_y & 1   & 0    &  0     \\
  0     &0     & 0     & 0    & 1   & 0    \\
  0     & 0     & 0    & 0    & -1/f_z   & 1
\end{array} \right) ,
\end{eqnarray}
where
\begin{eqnarray}
f_z=-\beta\gamma\cdot\frac{m_0 c^2}{2 \pi q}\frac{\beta^2\lambda sin\phi_s}{E_0 TL},
\end{eqnarray}

with  $\gamma$ and $\beta$ the lorentz factor and normalized velocity of the particle,
$\lambda$ the wavelength of the RF, $q$ the charge of the particle, and $E_0 T L$
the maximum energy gain or effective voltage of the cavity.

The transfer matrix of a system which is a combination of quadrupoles and drifts can be represented by
\begin{eqnarray}
M_{fodo}=\left( \begin{array}{cccccc}
 R_{11} & R_{12}  & 0       & 0        & 0    & 0  \\
R_{21}  & R_{22}  & 0       & 0        & 0    & 0   \\
  0     & 0       &R_{33}   & R_{34}   &0     &  0     \\
  0     & 0       &R_{43}   & R_{44}   & 0    &  0     \\
  0     &0        & 0       & 0        & 1    & R_{56}    \\
  0     & 0       & 0       & 0        & 0    &1
\end{array} \right) .
\end{eqnarray}
If the length of such a system is $L_d$, then $R_{56}=L_d/\gamma^2$.

The transfer matrix of a drift space with a length of $d$ at low energy can be written as
\begin{eqnarray}
M_{d}=\left( \begin{array}{cccccc}
 1      &  d       & 0   & 0     & 0    & 0  \\
 0      &  1       & 0   & 0     & 0    & 0   \\
  0     & 0        & 1   & d     &0     &  0     \\
  0     & 0        & 0   & 1     & 0    &  0     \\
  0     & 0        & 0    & 0    & 1    & d/\gamma^2    \\
  0     & 0        & 0    & 0    & 0    & 1
\end{array} \right).
\end{eqnarray}

The whole dog-leg section can be described by matrix multiplication:
\begin{eqnarray}
M=M_{b-} \cdot M_{D} \cdot M_{gap2} \cdot M_{fodo} \cdot M_{gap1}\cdot M_{d} \cdot M_{b_+},
\end{eqnarray}
where $M_{b-}$ and $M_{b_+}$ are the transfer matrices of bending magnets which bend the beam
to the left and right, $M_{D}$ and $M_{d}$ are the transfer matrices of
of drifts with length of $D$ and $d$, $M_{gap1}$ and $M_{gap2}$ are the transfer matrices of RF gaps with
longitudinal focal length of $f_z$ and $F_z$, $M_{fodo}$ is the transfer matrice
of a section composed of quadrupoles and drifts.

Assuming the two bending magnets are identical to each other, then $M_{b-}$ can be expressed as,
\begin{eqnarray}
M_{b_-}=\left( \begin{array}{cccccc}
 b_{11} & b_{12} & 0  & 0    & 0   & -b_{16}  \\
 b_{21} & b_{22} & 0  & 0    & 0   & -b_{26}   \\
  0     & 0      & 1  & b_{34}&0   &  0     \\
  0     & 0      & 0  & 1    & 0   &  0     \\
b_{26}  & b_{16}& 0  & 0    & 1   & b_{56}  \\
  0     & 0      & 0  & 0    & 0   & 1
\end{array} \right) .
\end{eqnarray}

After some tedious but straight forward matrix multiplication, we have:
\begin{eqnarray}
\label{eq7}
M_{15}=M_{62}=-b_{16}\cdot \left[ -\left( 1-R_{56}/F_z\right)/f_z -1/F_z\right],
\end{eqnarray}
and also,
\begin{eqnarray}
\label{eq8}
M_{25}=M_{61}=-b_{26}\cdot \left[ -\left( 1-R_{56}/F_z\right)/f_z -1/F_z\right],
\end{eqnarray}
where $M_{ij}$ is the element of matrix $M$ in the $i$th row and $j$th column.

To decouple the transverse from the longitudinal plane, it is required that
\begin{eqnarray}
 -\left( 1-R_{56}/F_z\right)/f_z -1/F_z=0 ,
\end{eqnarray}
or
\begin{eqnarray}
\label{eq9}
 R_{56}=f_z+ F_z.
\end{eqnarray}
We can see from Eq.~\ref{eq9} that $R_{56}$ or the distance between the two RF cavities
has a minimum when the two RF cavities have the same focusing length.

Given $f_z=F_z$, Eq.~\ref{eq9} can be written as,
\begin{eqnarray}
\label{eq10}
R_{56}=2f_z.
\end{eqnarray}

Assume  the synchronous phase of the two RF cavities are both
$-90$ degrees to maximize the bunching effect, Eq.~\ref{eq10} can be written as,
\begin{eqnarray}
\label{eq13}
 L_d=\frac{\lambda}{\pi}\cdot\frac{m_0 c^2}{q}\cdot\frac{\beta^3\gamma^3}{E_0 T L}.
\end{eqnarray}
This relation in Eq.~\ref{eq13} states that the distance of the two RF cavities is
uniquely determined by the their effective voltage to be able to
decouple the transverse from longitudinal. For high energy electron beams,
$\gamma\gg1$ gives $L_d\gg1$, which implies that no longitudinal bunching will
be necessary when a high energy electron beam is transporting a bending
section, and agrees with our common knowledge.

The other coupling terms, e.g. $M_{16}$ and $M_{26}$ have strong dependence
on the transverse focusing strengths or elements of $M_{fodo}$. The full decoupling
of transverse and longitudinal can
be achieved by adjusting the quadrupole strengths with the
help of simulation codes, such as TraceWin.

It is worth mentioning that a similar analysis can be carried out for a bending section with the
presence of only one bunching cavity. It can be easily found that the uncoupled achromatic condition
could not be fulfilled, thus at least two bunching cavities will be needed for such a dog-legsystem.

\section{Simulation result}
The input beam parameters are assumed to be the same as the ones for C-ADS MEBT2
except the beam current is set to zero in order to eliminate space charge effect.
The detailed beam parameters used for simulations are summarized in Table~\ref{beampara}.
\begin{center}
\tabcaption{ \label{beampara}  Input beam parameters used for simulation of the C-ADS like dog-leg section.}
\footnotesize
\begin{tabular*}{80mm}{c@{\extracolsep{\fill}}ccc}
\toprule Parameter &  Value   & Unit \\
\hline
Current       & 0          & mA \\
Beam energy   & 10.0       & MeV \\
$\epsilon_x$  & 0.21       & mm$\cdot$mard \\
$\epsilon_y$  & 0.20       & mm$\cdot$mard \\
$\epsilon_z$  & 0.16       & mm$\cdot$mard \\
$\alpha_x$    & -1.44      &- \\
$\beta_x$     & 2.70       &  mm/($\pi \cdot$ mrad)\\
$\alpha_y$    & -1.42      & - \\
$\beta_y$     & 2.68       & mm/($\pi \cdot$ mrad)\\
$\alpha_z$    & -0.31      & - \\
$\beta_z$     & 0.99       & mm/($\pi \cdot$ mrad)\\
\bottomrule
\end{tabular*}
\end{center}

We check the validity of Eq.~\ref{eq13} with an C-ADS MEBT2 like dog-leg section. The dog-leg is composed of
two bending magnets, two sets of triplets and two bunching cavities.
The layout of the dog-leg section is shown in Fig.~{layout}.
\begin{center}
\includegraphics[width=8cm]{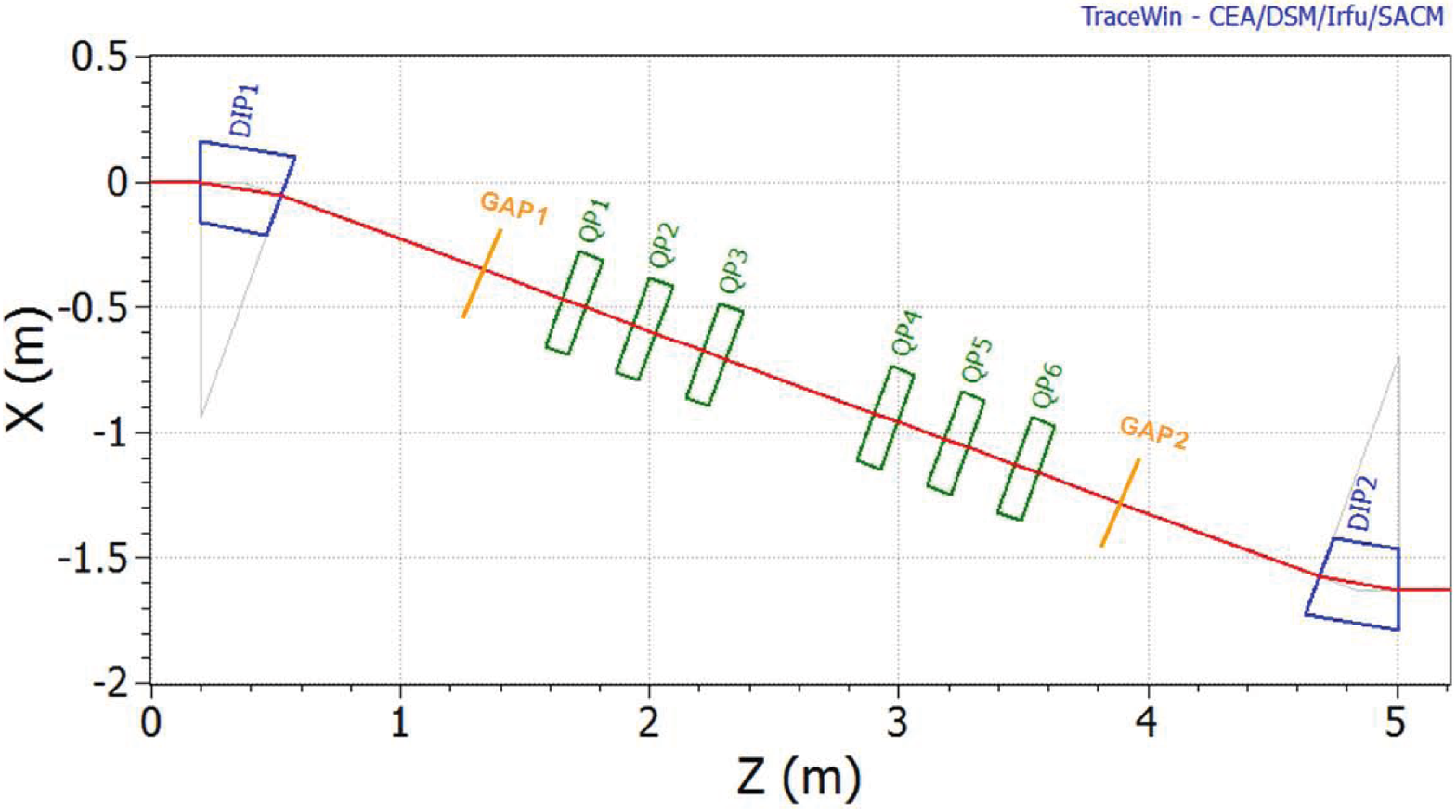}
\figcaption{\label{layout}   The layout of a C-ADS MEBT2 like dog-leg section.}
\end{center}

Two bending magnets are used in the dog-leg section and the two bending magnets are identical
with the exception that the first one bends the beam to the right while the second one bends the beam to the left.
The bending radius is $0.936$~m and the bending angle is $20$ degrees.
The bunching cavities are represented by
RF gaps which have synchronous phases of $-90$ degrees. The maximum energy gain of the RF gaps are assumed
to be $300$~kV, which is about twice the feasibility of normal conducting cavities working at continuous wave mode
at a beam energy of $10$~MeV. The distance between the two RF cavities is calculated by Eq.~\ref{eq13} and
been set up accordingly. The strength of the quadrupoles are been adjusted to fulfil
the achromatic condition, namely $R_{16}=R_{26}=0$. The detailed parameters list
can be found in Table~\ref{paralist}.
\begin{center}
\tabcaption{ \label{paralist}  Parameters list for the C-ADS like dog-leg section.}
\footnotesize
\begin{tabular*}{80mm}{c@{\extracolsep{\fill}}ccc}
\toprule Element & Length /mm   & Strength /T/m (kV) \\
\hline
D1 & 200 & - \\
B  & 326.8 & - \\
D2 & 800 & - \\
GAP1 & 0 &  300\\
D2 & 400 & - \\
Q1  & 100 & -7.2  \\
D3 & 200 & - \\
Q2  & 100 &  9.1 \\
D4 & 200 & - \\
Q3  & 100 &  -2.8 \\
D5 & 690 & - \\
Q4  & 100 & -2.8  \\
D6 & 200 & - \\
Q5  & 100 &  9.8 \\
D7 & 200 & - \\
Q6  & 100 & -8.5  \\
D8 & 400 & - \\
GAP2  & 0 &  300 \\
D9 & 800 & - \\
B  & 326.8 & - \\
D10 & 200 & - \\
\bottomrule
\end{tabular*}
\end{center}

The resulted transfer matrix elements $M_{15}$ ,$M_{62}$ and $M_{25}$ ,$M_{61}$
are shown in Fig.~\ref{r15} and Fig.~\ref{r25}, respectively.
\begin{center}
\includegraphics[width=8cm]{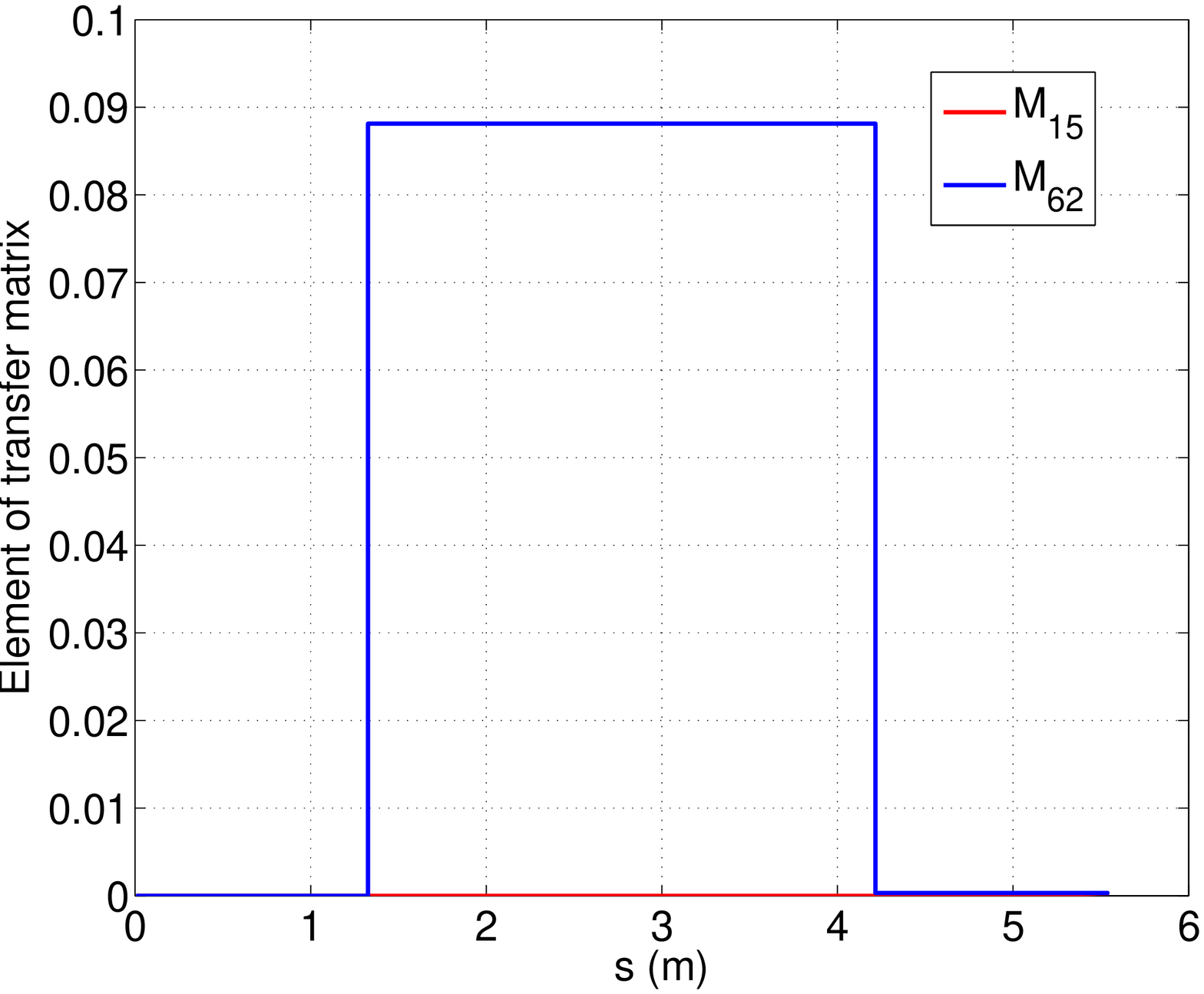}
\figcaption{\label{r15}   The $M_{15}$ and  $M_{62}$ elements of beam transfer matrix $M$.
These elements are predicted to be zero by Eq.~\ref{eq7}.}
\end{center}

\begin{center}
\includegraphics[width=8cm]{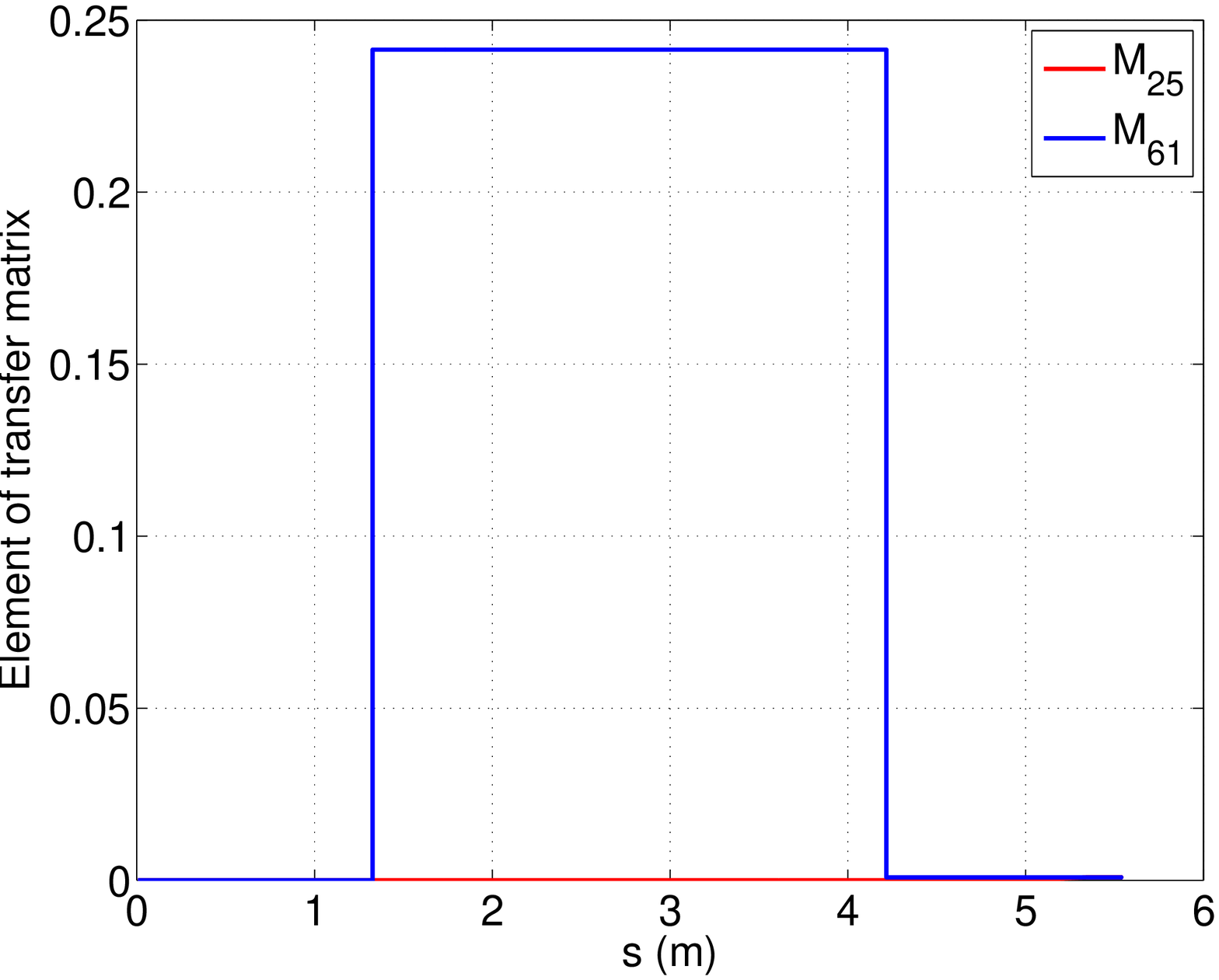}
\figcaption{\label{r25}   The $M_{25}$ and $M_{61}$ elements of beam transfer matrix $M$.
These elements are predicted to be zero by Eq.~\ref{eq8}.}
\end{center}
We can see from Fig.~\ref{r15} that $M_{61}$ and $M_{62}$ are produced by
the coupling of transverse to longitudinal when passing through the RF gap at a dispersive section.
They can be eliminated after passing through the second gap by properly choosing the distance between the two RF gaps.
$M_{15}$ and $M_{25}$ maintains zero till the second bending magnet. The final value of
$M_{15}$ and $M_{25}$ at the exit of the second bending magnet can be suppressed by
properly choosing the distance between the two RF gaps as shown in Eq.~\ref{eq7} and Eq.~\ref{eq8}.
It should be reminded that these properties of $M_{61}$, $M_{62}$, $M_{15}$ and $M_{25}$ are not subject
to the strength of the qudrupoles along the bending section.

\begin{center}
\includegraphics[width=8cm]{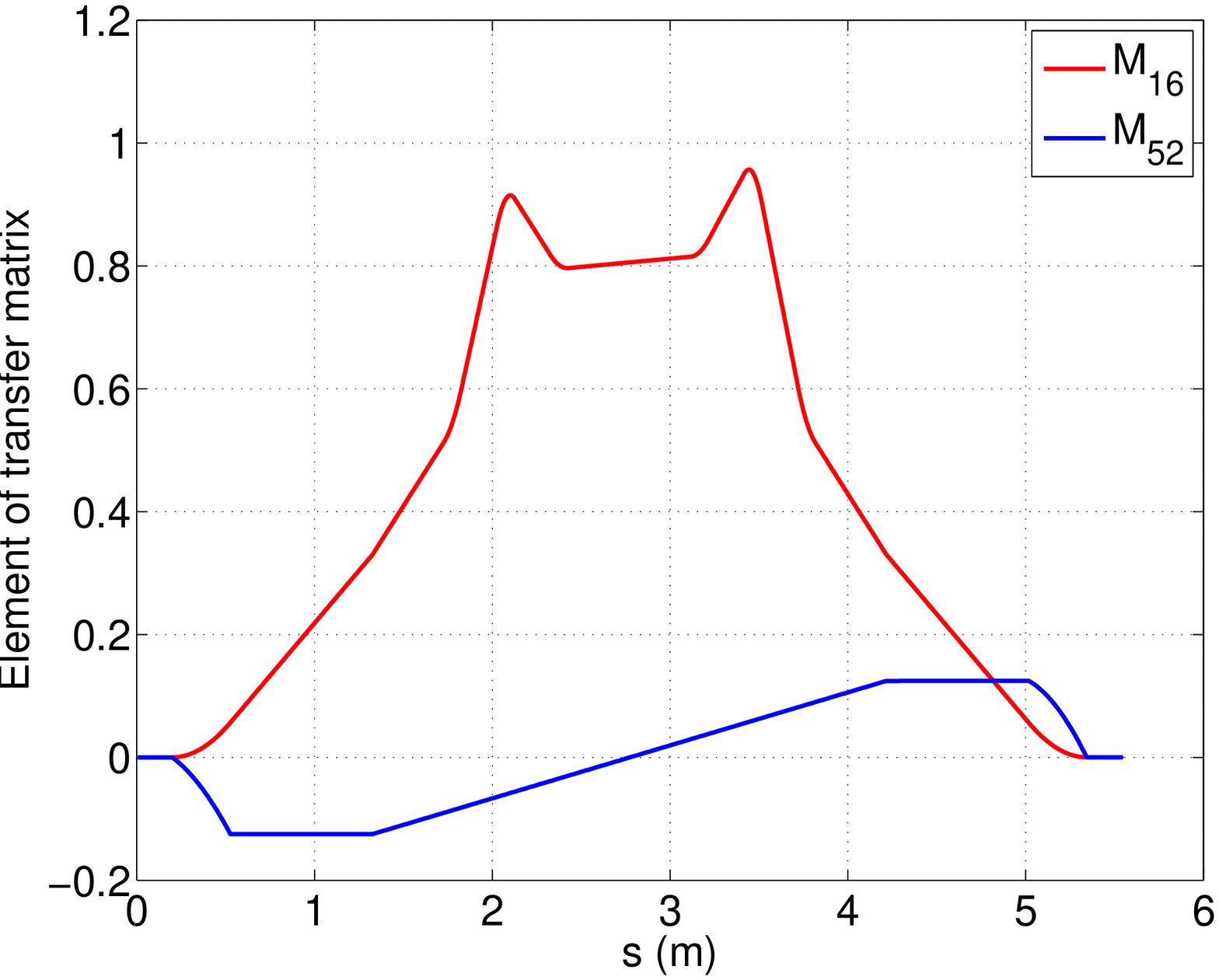}
\figcaption{\label{r16}  The $M_{16}$ and $M_{51}$ elements of beam transfer matrix.
The strengths of quarupoles are adjusted to have $M_{16}=M_{26}=0$.}
\end{center}
The dispersive terms $M_{16}$ and $M_{26}$ are strongly depend on the setup of the quadrupoles strengths. The
achromatic condition can be fulfilled by properly adjusting the strengths of the six quadupoles. The results
are shown in Fig.~\ref{r16} and Fig.~\ref{r26}.
We can see that both $M_{16}$ , $M_{26}$ and $M_{51}$, $M_{52}$ and returns to zero after exiting the second bend.
\begin{center}
\includegraphics[width=8cm]{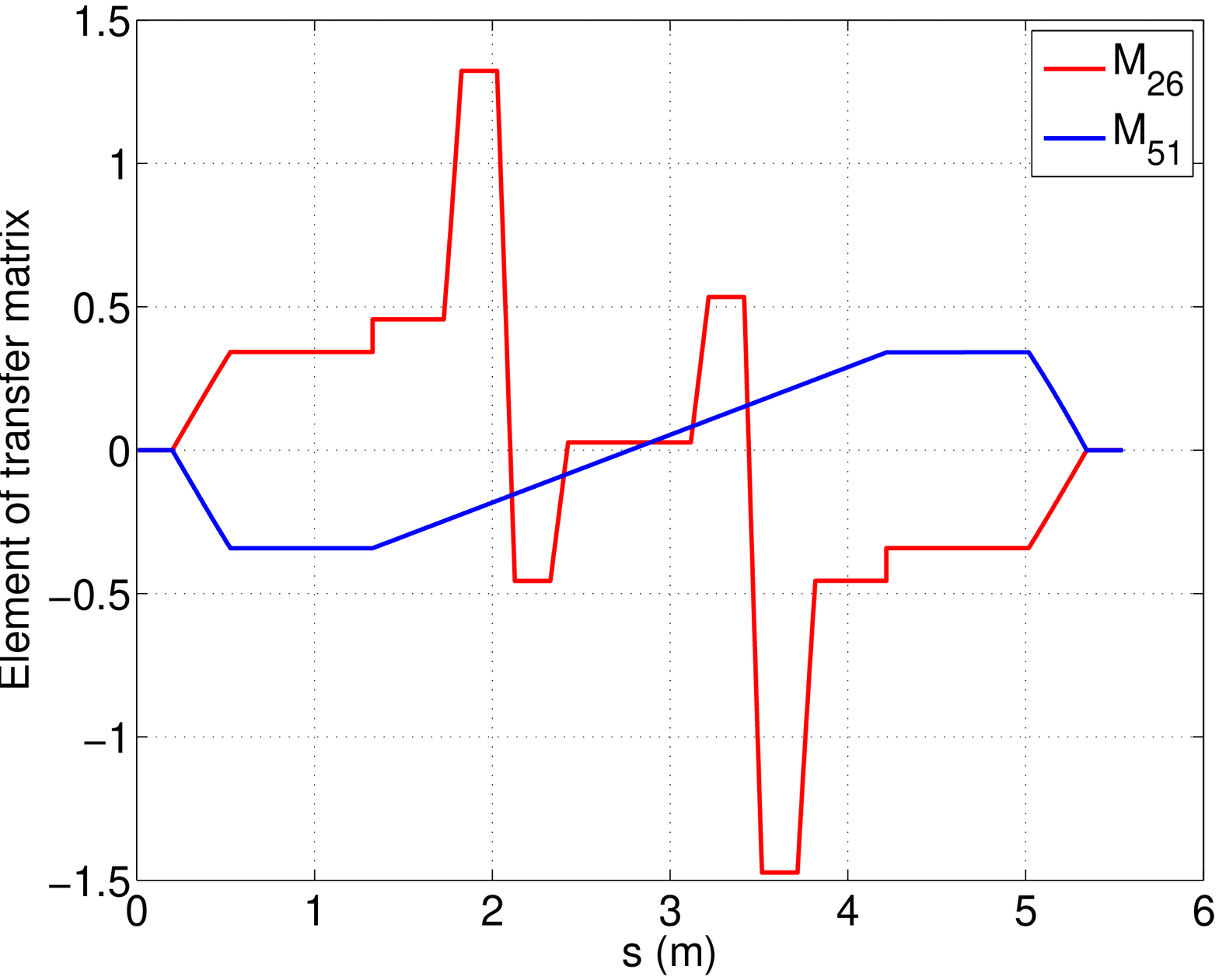}
\figcaption{\label{r26}  The $M_{26}$ and $M_{51}$ elements of beam transfer matrix.
The strengths of quarupoles are adjusted to have $M_{16}=M_{26}=0$.}
\end{center}

The relationship of Eq.~\ref{eq10} also guarantees another important feature of the dog-leg system,
which is the reversal of the longitudinal phase space, namely $M_{55}=M_{66}=-1$.
The reversal of the longitudinal phase space also explains why the dog-leg system is
symmetric instead of anti-symmetric achromatic. The $M_{55}=M_{66}=-1$ of the dog-leg system
is shown in Fig.~\ref{r55}.
\begin{center}
\includegraphics[width=8cm]{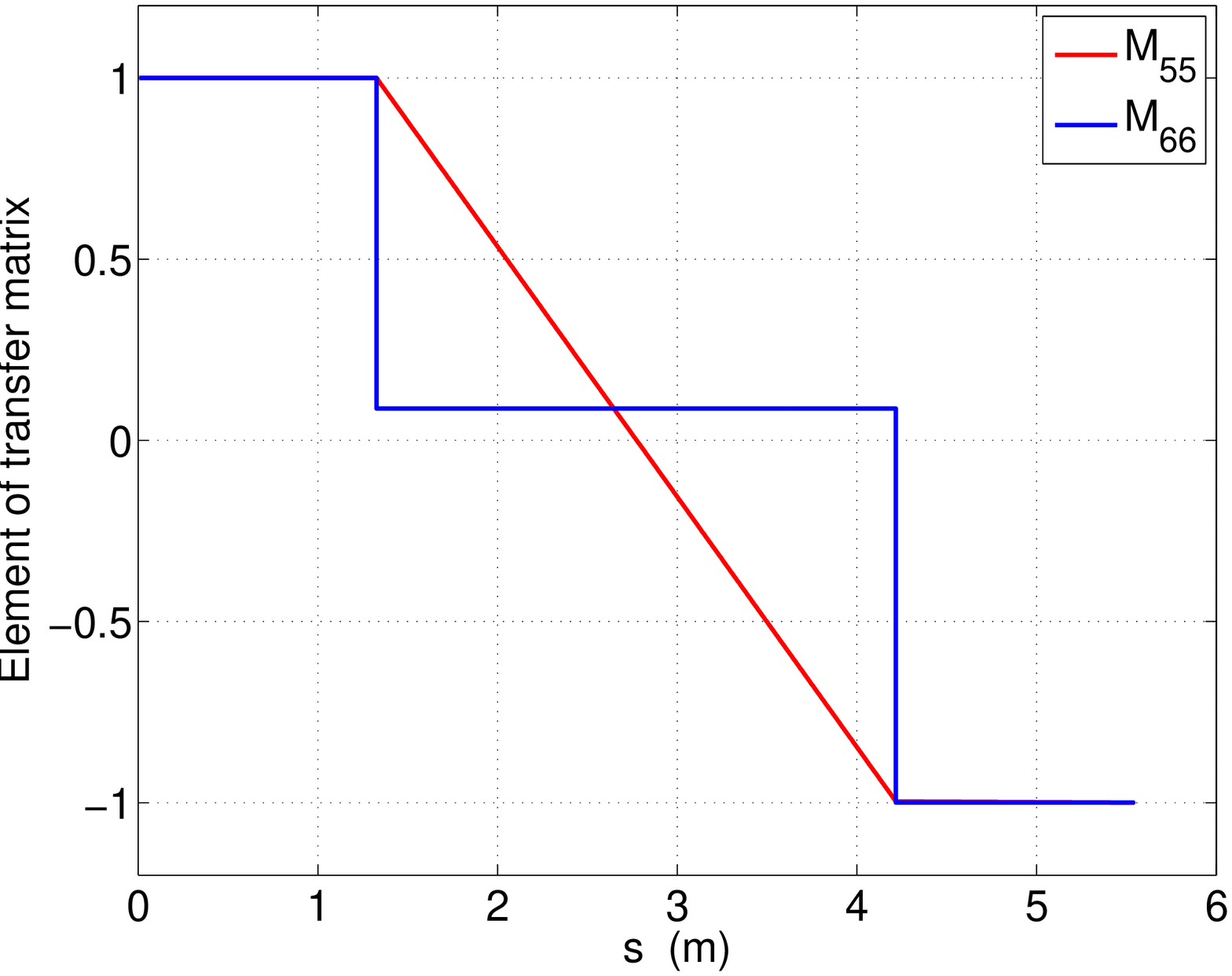}
\figcaption{\label{r55}  The $M_{55}$ and $M_{66}$ elements of beam transfer matrix.
$M_{55}= M_{66}=-1$ or the reversal of the longitudinal phase space is guaranteed by Eq.~\ref{eq10}. }
\end{center}

Since the whole system is fully uncoupled as can be seen in Fig.~\ref{r15} to Fig.~\ref{r26}, we would expect no growth
of projected rms emittance in all three planes. The resulted projected emittance of the dog-leg system are shown in Fig.~\ref{rmsemit}.
It is easy to see that the rms emittances remain the same after passing through the whole system. The growth of projected emittance in
the middle of the bending section is caused by the coupling of transverse and longitudinal planes.
\begin{center}
\includegraphics[width=8cm]{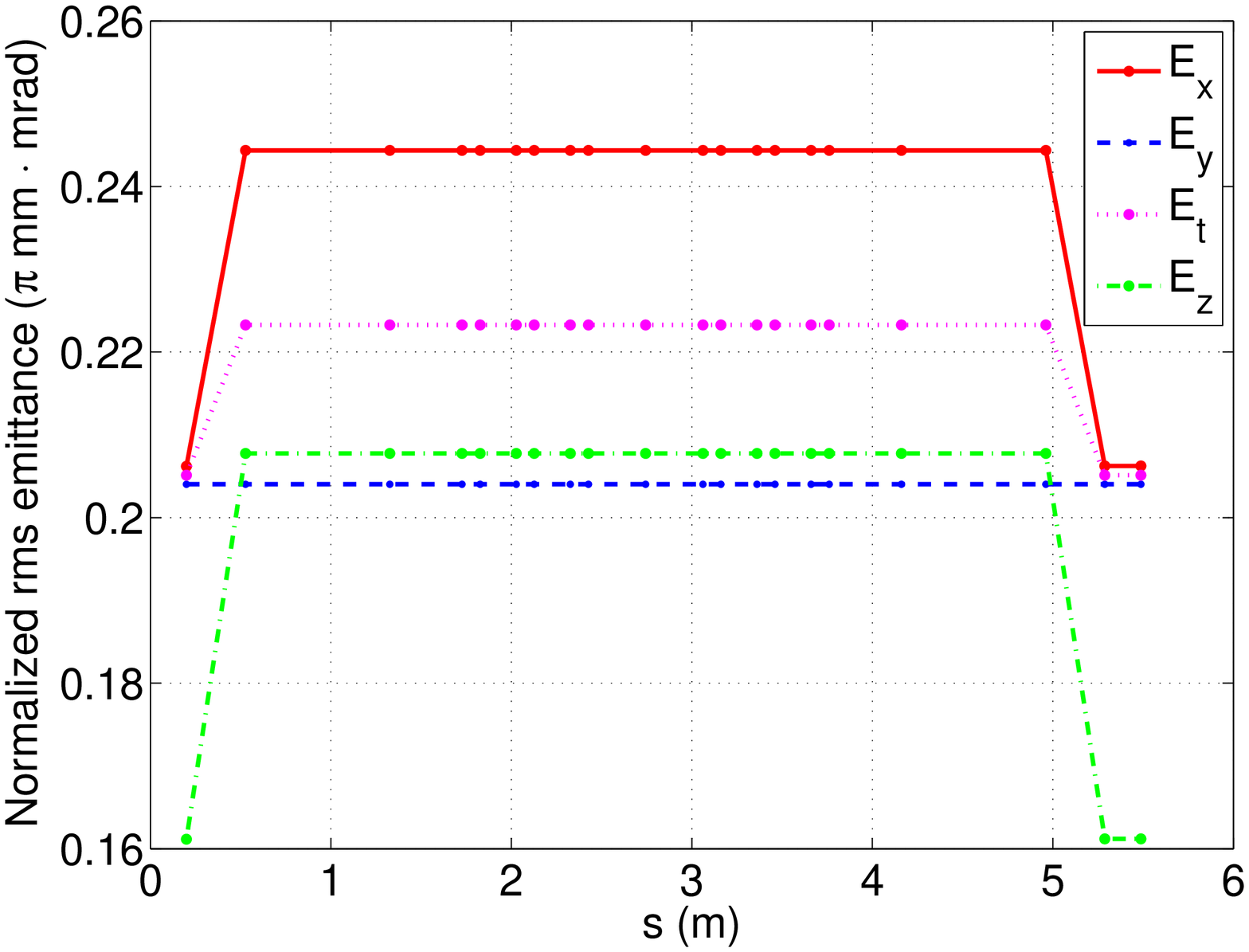}
\figcaption{\label{rmsemit} The evolution of projected rms emittances in $x$,$y$, $4$D transverse and $z$ planes.}
\end{center}

\section{Space charge effect}
The inclusion of space charge effect will break the simple law of Eq.\ref{eq13} to fully decouple transverse from
longidudinal.  The distortion of uncoupling condition  from space charge effect will lead
to the growth of the projected emittance. It has been proved that decoupling with the exclusion of space charge effect is an essential way
to minimize the beam emittance growth~\cite{guozhen_thesis}.
We simulate the effect of space charge effect on the emittance growth of the dog-leg section described in
the previous section. Gaussian distributions in all three planes are assumed. Ten thousand macro particles are
used in the simulation. We keep the input beam parameter the same and gradually increase beam current
from $0$~mA to $40$~mA. The change of projected rms emittances at the end of the dog-leg section
in $x$,$y$ and $z$ planes as the beam current increases
are shown in Fig.~\ref{current}.

\begin{center}
\includegraphics[width=8cm]{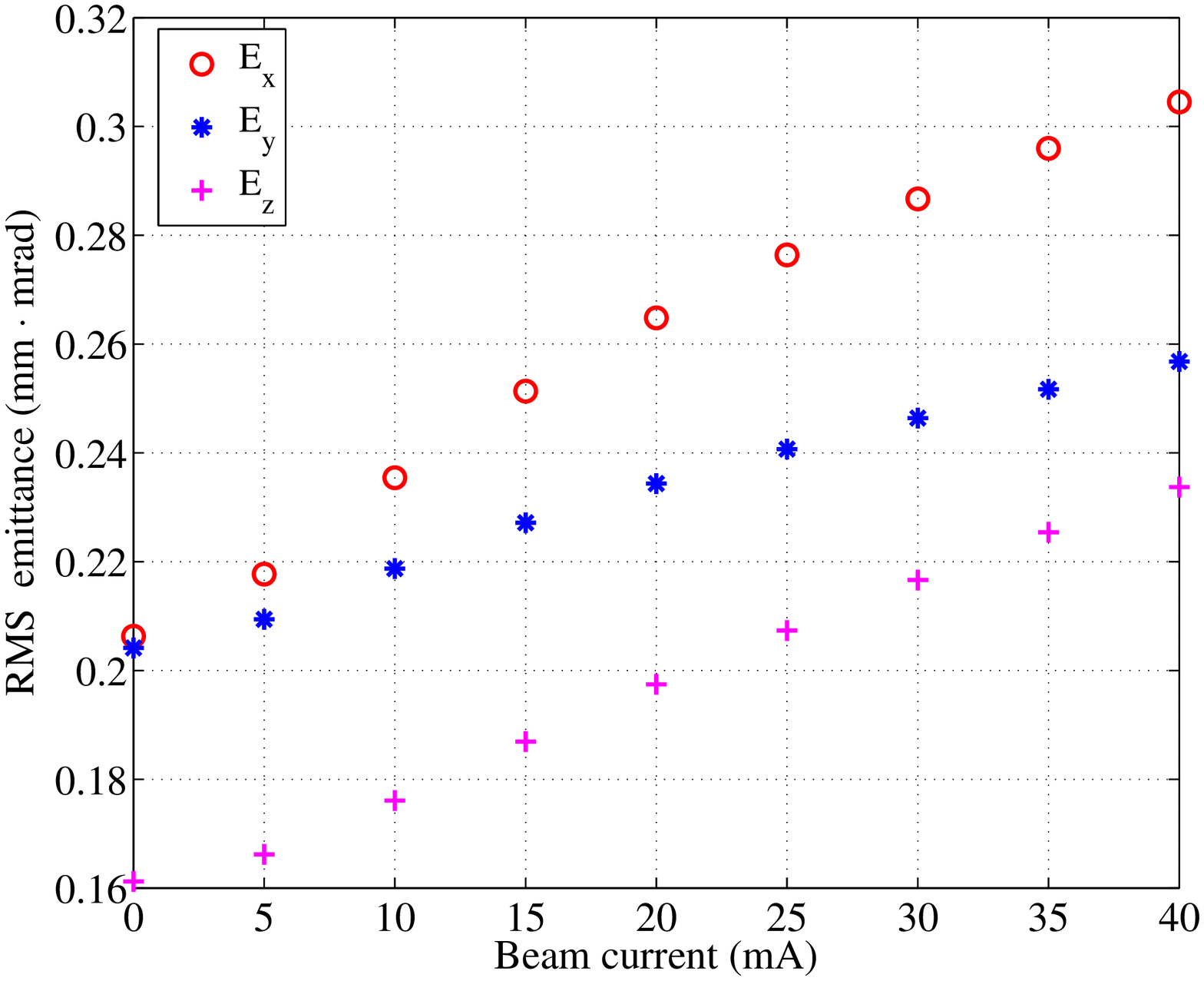}
\figcaption{\label{current} The change of projected rms emittances in $x$,$y$ and $z$ planes as the beam current increases
from $0$~mA to $40$~mA.}
\end{center}

We can see that the projected emittance grows in all three planes as the beam current increases.
The growth rate of the emittances in $x$ and $z$ planes are consistent with each other
due to the residual coupling caused by space charge effect. The emittance in $y$ plane grows slower
than in the other two planes since the coupling is weaker with the absence of bending in $y$ plane.

\section{Conclusions}

The uncoupled achromatic condition have been analyzed for a C-ADS like dog-leg system
with the presence of RF cavities. Theoretical analysis shown that at least two cavities
are required to decouple transverse from longitudinal planes. An explicit formula was
given to determine the distance between two cavities in order to realize the uncoupling.
It is shown that the distance between two cavities is inversely proportional to the
maximum energy gain in each cavity. Proof of principle simulations were done with the code TraceWin.
The simulation results agree well with the theoretical analysis.

The formula deduced in this paper can help determine the elements layout when designing a
dog-leg system which has to incorporate RF cavities.

\section{Acknowledgment}
\acknowledgments{Huiping Geng would like to thank Dr. Xiaoyu Wu and
Dr. P.A. Phi Nghiem for useful discussions.}

\end{multicols}

\clearpage

\end{CJK*}
\end{document}